%% file: manuscript.tex
\documentclass[jcp,aip,reprint,superscriptaddress,showpacs,floats,floatfix,twocolumns]{revtex4-1}
\pdfoutput=1
\usepackage[english]{babel}
\usepackage{color}
\usepackage{graphicx}
\usepackage{placeins}
\usepackage[caption=false]{subfig} 
\usepackage{amsmath,amssymb,bm}
\usepackage[version=3]{mhchem}
\usepackage{verbatim}
\usepackage{multirow}
\usepackage{dcolumn}
\usepackage{float}
\usepackage{nicefrac}
\usepackage{siunitx}
\usepackage{tikz}
\usepackage{pgfplots}
\usepackage{shellesc}
\usepackage{array}
\usepackage{booktabs}
\usepgfplotslibrary{groupplots}
\usepgfplotslibrary{colormaps}
\usepgfplotslibrary{fillbetween}
   \pgfplotsset{compat=1.11}
   \usetikzlibrary{backgrounds}
   \usetikzlibrary{shapes,snakes}
   \usetikzlibrary{calc}
   \usetikzlibrary{decorations.pathmorphing}
   \usetikzlibrary{positioning,arrows.meta,bending}
   \usepgfplotslibrary{external}
   \tikzset{external/figure name=tmpplot/figure, external/optimize command
away=\titlepage{0}}
\DeclareSIUnit[number-unit-product = {\,}]{\cal}{cal}
\DeclareSIUnit[number-unit-product = {\,}]{\kcal}{\kilo\cal}
\DeclareSIUnit[number-unit-product = {\,}]{\atomicunit}{a.u.}
\usepackage[colorlinks,allcolors=black,citecolor=blue,urlcolor=blue]{hyperref}

\definecolor{rubgrey}{rgb}{0.90,0.89,0.89}
\definecolor{rubgreen}{rgb}{0.58,0.76,0.11}
\definecolor{rubblue}{rgb}{0.0,0.21,0.38}
\definecolor{rubdarkgrey}{rgb}{0.60,0.59,0.59}
\definecolor{rubwhite}{rgb}{1.0,1.0,1.0}
\definecolor{fugrey}{rgb}{0.95,0.95,0.95}
\definecolor{resolvblue}{rgb}{0.00,0.925,1.00}
\definecolor{fzjblue50}{RGB}{0,91,130}
\definecolor{fzjblue35}{RGB}{166,198,211}
\definecolor{fzjblue30}{RGB}{178,206,217}
\definecolor{fzjblue20}{RGB}{204,222,236}
\definecolor{fzjblue10}{RGB}{229,239,242}
\definecolor{fzjgray80}{RGB}{81,81,81}
\definecolor{fzjgray50}{RGB}{156,156,156}
\definecolor{fzjgray30}{RGB}{185,185,185}
\definecolor{fzjgray20}{RGB}{204,204,204}
\definecolor{fzjgray10}{RGB}{229,229,229}
\definecolor{fzjgray05}{RGB}{242,242,242}
\definecolor{fzjwhite}{RGB}{255,255,255}
\definecolor{fzjbranchred}{RGB}{212,45,18}% Ges
\definecolor{fzjbranchyellow}{RGB}{230,175,17}%
\definecolor{fzjbranchblue}{RGB}{0,131,190}% In
\definecolor{fzjblue}{RGB}{0,91,130}% Diagra
\definecolor{fzjlightblue}{RGB}{110,159,189}
\definecolor{fzjred}{RGB}{175,90,80}% Diagram
\definecolor{fzjlightred}{RGB}{198,141,132}% D
\definecolor{fzjgreen}{RGB}{125,150,110}% Diag
\definecolor{fzjlightgreen}{RGB}{164,181,153}%
\definecolor{fzjyellow}{RGB}{215,170,80}% Diag
\definecolor{fzjlightyellow}{RGB}{235,212,167}
\definecolor{molRed}{RGB}{192, 20, 45}
\definecolor{HeBlue}{RGB}{95, 80, 130}

\newcommand{\forloop}[5]{%
   \setcounter{#1}{#2}%
   \ifthenelse{#2 < #4}{%
      \whiledo{\value{#1} < #4}{%
         #5%
         \addtocounter{#1}{#3}%
      }%
      \ifthenelse{\value{#1} = #4}{%
         #5%
      }{}%
   }{%
      \ifthenelse{#2 = #4}{%
         #5%
      }{}%
   }%
}%
\newread\outputstream
\newcommand{\inputtikz}[1]{%
   \makeatletter %
   \filename@parse{#1} %
   \makeatother

   \tikzsetnextfilename{\filename@area\filename@base}%
   \input{#1}%
}

\DeclareSIUnit{\calorie}{cal}
\DeclareSIUnit{\rydberg}{Ry}
\DeclareSIUnit{\atomicunit}{a.u.}
\DeclareSIUnit{\electrons}{\ensuremath{e}}
\DeclareSIUnit{\wavenumber}{\per\centi\metre}
\DeclareSIUnit{\arbu}{arb.~units}
\DeclareSIPostPower\tothefourth{4}
\DeclareSIPostPower\tothefifth{5}
\DeclareSIPostPower\tothesix{6}

\newcommand{\zundel}{\cf{H5O2+}}%

\begin{document}
\title{%
    Automated Fitting of Neural Network Potentials
at Coupled Cluster Accuracy:
Protonated Water Clusters as Testing Ground
}

\author{Christoph Schran}
\email{Christoph.Schran@rub.de}
\affiliation{Lehrstuhl f\"ur Theoretische Chemie,
  Ruhr--Universit\"at Bochum, 44780 Bochum, Germany}
\author{J\"org Behler}
\affiliation{Universit\"at G\"ottingen,
    Institut f\"ur Physikalische Chemie,
    Theoretische Chemie,
    Tammannstr. 6, 37077 G\"ottingen, Germany}
\author{Dominik Marx}
\affiliation{Lehrstuhl f\"ur Theoretische Chemie,
  Ruhr--Universit\"at Bochum, 44780 Bochum, Germany}
\date{\today}

\keywords{
Potential Energy Surface,
Neural Network Potentials,
Protonated Water Clusters
}

\begin{abstract}
    Highly accurate potential energy surfaces are of key
    interest for the detailed understanding and predictive modeling
    of chemical systems.
    In recent years, several new types of force fields,
    which are based on machine learning algorithms and
    fitted to \textit{ab initio} reference calculations,
    have been introduced to meet this requirement.
    Here we show how high--dimensional neural network potentials
    can be 
    employed
    to automatically generate the potential
    energy surface of finite sized clusters at 
    coupled cluster accuracy, 
namely CCSD(T*)-F12a/aug-cc-pVTZ.
    The developed automated procedure utilizes the established intrinsic properties
    of the model such that the configurations for the
    training set are selected in an unbiased and efficient way
    to minimize the computational effort of expensive reference calculations.
    These ideas are applied to protonated water clusters 
    from the hydronium cation, \cf{H3O+}, up
    to the tetramer, \cf{H9O4+}, and 
lead to a 
single
potential
    energy surface 
that describes all these systems
    at essentially converged coupled cluster
    accuracy with a fitting error of 0.06~kJ/mol per atom.
    The fit is validated in detail 
    for all clusters up to
    the tetramer and yields reliable results not
    only for stationary points, but also for reaction
    pathways, intermediate configurations, as well as
    different sampling techniques.
    Per design the NNPs constructed in this fashion can handle
    very different conditions including the quantum nature
    of the nuclei and enhanced sampling techniques 
covering 
    very low as well as high temperatures.
    This enables fast and exhaustive exploration
    of the targeted protonated water clusters
with
essentially converged interactions.
    In addition, the automated process will
    allow 
one
    to tackle 
finite systems 
    much beyond the present case.
\end{abstract}

\maketitle

\section{Introduction}
\label{sec:intro}

The potential energy surface 
(PES)
of a system, which governs all
its structural, dynamic and thermodynamic properties in the Born--Oppenheimer approximation, is currently
in many cases
best described
by coupled cluster theory,
granting 
the CCDS(T) approach 
the title of the ``gold standard''
in quantum chemistry~\cite{Bartlett2007/10.1103/RevModPhys.79.291,
Hobza2012/10.1021/ar200255p}.
Thus, it is desirable to utilize this method 
not only in the realm of static single--point calculations, but also for finite temperature dynamical simulations 
in order to reach chemical accuracy and provide predictive
answers purely based on simulations.
However, the excellent quality comes at a rather high price in terms
of the computational cost that, moreover, scales very unfavorably
with the system size.
Advanced simulations, such as rare event sampling techniques or
path integral methods, that typically require many million evaluations
of the interactions to reach convergence are, therefore, usually out of scope if
energies and forces 
need to be 
evaluated ``on--the--fly''~\cite{Marx2009} during the simulation. 
This is the very reason why only notable exceptions 
exist~\cite{Spura2015/10.1039/C4CP05192K,
Mouhat2017/10.1021/acs.jctc.7b00017,
Haycraft2017/10.1021/acs.jctc.6b01107}
where correlated electronic structure methods such as 
coupled cluster theory have been applied on--the--fly to small
molecules such as the Zundel cation, \zundel{}. 

Especially for the study of water, different
approaches based on physically motivated functional forms
have been very successful in reaching 
high accuracy
as for example shown for the MB-Pol water force
field~\cite{Babin2013/10.1021/ct400863t}
or other highly accurate fitting
schemes~\cite{Bukowski2007/10.1126/science.1136371,
Wang2009/10.1063/1.3196178,Babin2012/10.1021/jz3017733};
see e.g. Ref.~\citenum{Cisneros2016/10.1021/acs.chemrev.5b00644}
for more examples.
Being usually based on many--body expansions, such 
approaches were also recently applied to
protonated water clusters, explicitly including
up to 
four body terms~\cite{Heindel2018/10.1021/acs.jctc.8b00598},
and shown to 
reproduce coupled cluster reference
calculations with rather high 
accuracy
for stationary
point structures.
Other notable potentials for protonated water (clusters)
rely on empirical models~\cite{Kozack1992/10.1063/1.461957},
perturbation theory~\cite{Hodges1999/10.1063/1.478580}, or
are based on empirical valence bond models
with increasing complexity of the reference
states~\cite{Voth1996/10.1063/1.470962,
Schmitt1998/10.1021/jp9818131,
Vuilleumier1998/10.1016/S0009-2614(97)01365-1,
James2005/10.1063/1.1869987,
Brancato2005/10.1063/1.1902924,
Wu2008/10.1021/jp076658h,
Kumar2009/10.1021/jp8066475}.

At the same time, prominent advances in machine learning techniques
have led to the development of 
computationally very efficient, yet
accurate potential energy surfaces~\cite{Handley2010/10.1021/jp9105585,
Behler2011/10.1039/C1CP21668F,
Behler2016/10.1063/1.4966192,
Bartok2017/10.1126/sciadv.1701816,
Butler2018/10.1038/s41586-018-0337-2},
where various different techniques have been introduced
over the years~\cite{Blank1995/10.1063/1.469597,
Prudente1998/10.1063/1.477550,
Lorenz2004/10.1016/j.cplett.2004.07.076,
Lorenz2006/10.1103/PhysRevB.73.115431,
Manzhos2006/10.1021/jp055253z,
Behler2007/10.1103/PhysRevLett.98.146401,
Bartok2010/10.1103/PhysRevLett.104.136403,
Rupp2012/10.1103/PhysRevLett.108.058301,
Shapeev2015/10.1137/15M1054183,
Li2015/10.1103/PhysRevLett.114.096405,
Thompson2015/10.1016/j.jcp.2014.12.018,
Schuett2017/10.1038/ncomms13890,
Chmiela2017/10.1126/sciadv.1603015,
Faraji2017/10.1103/PhysRevB.95.104105,
Unke2019/10.1021/acs.jctc.9b00181}.
These methods do not utilize physically motivated functional
forms,
but rather use highly flexible general functions
being able to represent in principle arbitrary 
functional
relations.
They are usually trained to electronic structure reference data
and can afterwards reproduce the structure--energy relation
with
high
precision.
The first such technique, 
which is 
scalable to 
essentially
arbitrary system sizes and based on artificial neural networks, 
is the high--dimensional neural
network potential (NNP) methodology~\cite{Behler2007/10.1103/PhysRevLett.98.146401,
Behler2017/10.1002/anie.201703114}
that has been proven to be well suited for the description
of a variety of systems as summarized in
Ref.~\citenum{Behler2017/10.1002/anie.201703114}.
In addition, the high flexibility of the underlying functional relation
provides a very powerful tool for the identification of
deficiencies in the training set, as first
mentioned in Ref.~\citenum{Behler2011/10.1039/C1CP21668F}
and used in Ref.~\citenum{Artrith2012/10.1103/PhysRevB.85.045439}.
By comparison of two distinct machine learning potentials
which feature large differences for configurations
not sufficiently represented,
the training set can be iteratively optimized~\cite{Artrith2012/10.1103/PhysRevB.85.045439}.
This has enabled the development of
iterative strategies for the
assembling of the training set
for NNPs for example 
using the notion of
adaptive sampling~\cite{Gastegger2017/10.1039/c7sc02267k}.
In addition, similar strategies were
recently also applied 
to
other machine learning
approaches 
in the context of
Gaussian approximation potentials~\cite{Bartok2010/10.1103/PhysRevLett.104.136403}
called data driven learning~\cite{Deringer2018/10.1103/PhysRevLett.120.156001},
or for moment tensor potentials~\cite{Shapeev2015/10.1137/15M1054183}
called active learning~\cite{Podryabinkin2017/10.1016/j.commatsci.2017.08.031}.
While most of these machine learning approaches
have been targeting DFT (density functional theory) reference calculations,
recent studies employing machine
learning potentials were able to closely reproduce
coupled cluster accuracies for finite sized
molecular clusters~\cite{Schran2018/10.1063/1.4996819,Schran2018/10.1021/acs.jctc.8b00705,Chmiela2018/10.1038/s41467-018-06169-2,
Smith2019/10.1038/s41467-019-10827-4}.

In the following, we will present how high--dimensional neural network
potentials can be utilized for the automated fitting of highly--accurate
potential energy surfaces to reach coupled cluster accuracy.
In the present study, protonated water clusters 
from the monomer
up to the tetramer
are chosen as a case study for which converged coupled cluster reference
calculations are still feasible.
The NNP methodology has already been successfully applied to
these clusters using a density functional theory based
description for the reference calculations~\cite{Natarajan2015/10.1039/C4CP04751F}.
In the present work, rather similar fitting accuracies are reached, but the 
machine learning
methodology is used to reproduce
much more accurate so--called
CCSD(T*)-F12a/aug-cc-pVTZ
energies, 
which provide essentially converged coupled cluster data
as described in the computational details. 
As mentioned above,
it is important to note that for molecular systems 
such as water for instance, 
very accurate alternative approaches making use of N-body expansions 
have also been proposed.
Examples are as diverse as 
permutationally 
invariant polynomials
(PIPs)
\cite{Braams2009/10.1080/01442350903234923},
the MB-pol water potential~\cite{Babin2013/10.1021/ct400863t}
and also machine learning potentials incorporating low-order many-body
terms~\cite{Bartok2013/10.1103/PhysRevB.88.054104}.
These 
and related
approaches offer the advantage that they are likely to have a better
transferability and extrapolation performance for larger
systems
beyond those 
that have been explicitly covered 
by the underlying training set. 
On the other hand, 
advantages of our approach are its very general
applicability to 
many different types of 
systems also beyond those consisting of
fixed molecular entities, such as metal
clusters~\cite{Artrith2012/10.1103/PhysRevB.85.045439},
the 
global
permutation invariance 
of all chemically equivalent atoms in arbitrarily large systems, 
and the resulting 
unconstrained
reactivity 
as required for instance to describe long-range 
proton transport processes in bulk systems, 
which has been demonstrated already
for aqueous electolytes like bulk
NaOH(aq)~\cite{Hellstrom2016/10.1021/acs.jpclett.6b01448}
as well as for water dissociating at zinc oxide 
surfaces~\cite{Quaranta2017/10.1021/acs.jpclett.7b00358}.

The remaining manuscript is structured as follows:
First, the automated fitting procedure
for neural network potentials
is reviewed
in detail.
Afterwards, the quality of the resulting potential
energy surface is validated with respect
to the coupled cluster reference 
not only 
for stationary--point structures,
but 
also for reaction pathways and
intermediate configurations.
Finally, the performance of the NNP
for molecular dynamics
and path integral simulations
is verified explicitly.
\section{Automated Fitting Procedure for Neural Network Potentials}
\label{sec:methods}

\begin{figure}[t]
    \centering{}
    \includegraphics[width=1.0\linewidth]{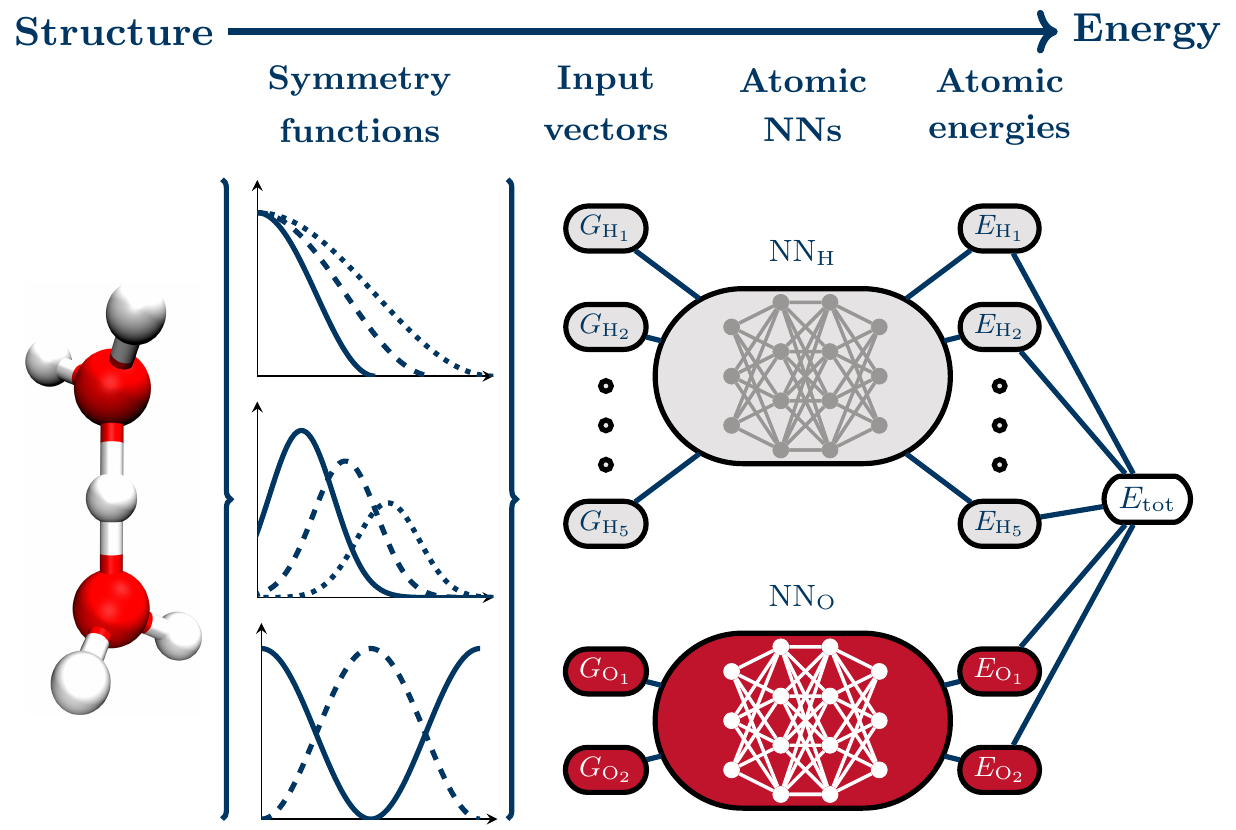}
    \caption{
        Representation of the structure--energy relation
        in a high--dimensional neural network potential
(NNP) 
        for the specific example of a Zundel cation \zundel{}.
        In a first step, the structure is transformed via
        atom--centered symmetry functions into rotationally and
        translationally invariant vectors
        each of which is represented by a small box $G$.
        These serve as the input for atomic neural networks
        to provide atomic energies that
        sum up to the total energy of the system.
        This functional relation is analytically
        differentiable and, thus, can provide the forces
        of the system.
    }
    \label{fig:hdnnp}
\end{figure}

Application of neural networks to chemically complex systems
can be achieved, as reviewed in detail in
Ref.~\citenum{Behler2017/10.1002/anie.201703114}, by
the construction of a structure--energy relation via a transformation
of the structure by atom--centered 
so--called
symmetry functions that provide the input for atomic neural networks.
These neural networks output atomic energy contributions
for all atoms in the system that sum up to the total energy. 
The resulting functional relation, as schematically depicted in Fig.~\ref{fig:hdnnp}
for the Zundel cation \zundel{}, can be analytically differentiated
to provide the forces of the system.
For further details on atom--centered symmetry functions and the
high--dimensional neural network approach we refer the
reader to Refs.~\citenum{Behler2011/10.1063/1.3553717}
and~\citenum{Behler2017/10.1002/anie.201703114}, respectively,
as well as
to the supporting information 
where our particular approach is explained in depth.
and recall only that the resulting NNPs are also permutationally invariant.

For the purpose of this study, it is sufficient to
stress that this approach provides a highly flexible
functional form that can be fitted to a training set
with configurations for which the corresponding energies are known.
This fit can, afterwards, be used for accurate
interpolation between the training points, while extrapolation
outside of the configuration space spanned by the training
set usually becomes unreliable due to the absence of physically
motivated functional forms.
A key task 
in
the development of neural network potentials is,
therefore, the preparation of the reference set used for the
training of the potential.
It needs to be representative
for
all envisaged conditions and 
must
sample the spanned configuration space in a balanced manner.

The intrinsic properties of the neural network approach
provide a very powerful tool for this difficult task, as first
introduced in Ref.~\citenum{Behler2011/10.1039/C1CP21668F}
and used in Ref.~\citenum{Artrith2012/10.1103/PhysRevB.85.045439}.
The high flexibility of the functional form results
usually in large differences for regions in configuration
space that are underrepresented in the training set.
Using this approach, unnecessarily large numbers of reference calculations
can be avoided, since the calculations can be restricted
to the
important configurations really needed
for an
improvement of the potential
only where actually required. 
We note in passing 
that this strategy is well known in
the field of active learning 
as the ``query by committe'' approach~\cite{Seung1992/10.1145/130385.130417}.
In addition, it has been proven to feature
an error that decreases exponentially
with the number of training points
which already holds for the smallest two-member
committee~\cite{Seung1992/10.1145/130385.130417}
that we use here.
A second strategy to identify configurations
for an improvement of the training set relies on the identification
of configurations beyond the boundaries of
the configuration space spanned by the training set.
Such points are likely to show extrapolation errors
and can be identified by comparing the description
of the structure encoded in the symmetry function values
to the range of the symmetry function
values encountered in the training set.
Methods to select the most important configurations
for an improvement of the training set
are especially important if 
high--level
and thus demanding
quantum chemistry methods such as 
CCSD(T)
theory shall be used.
It also allows for a high level of automation of the
development 
and the systematic improvement
of potential energy surfaces
relying on as few as possible reference calculations
as explained in detail in the following.

\begin{figure}[t]
    \centering{}
    \includegraphics[width=1.0\linewidth]{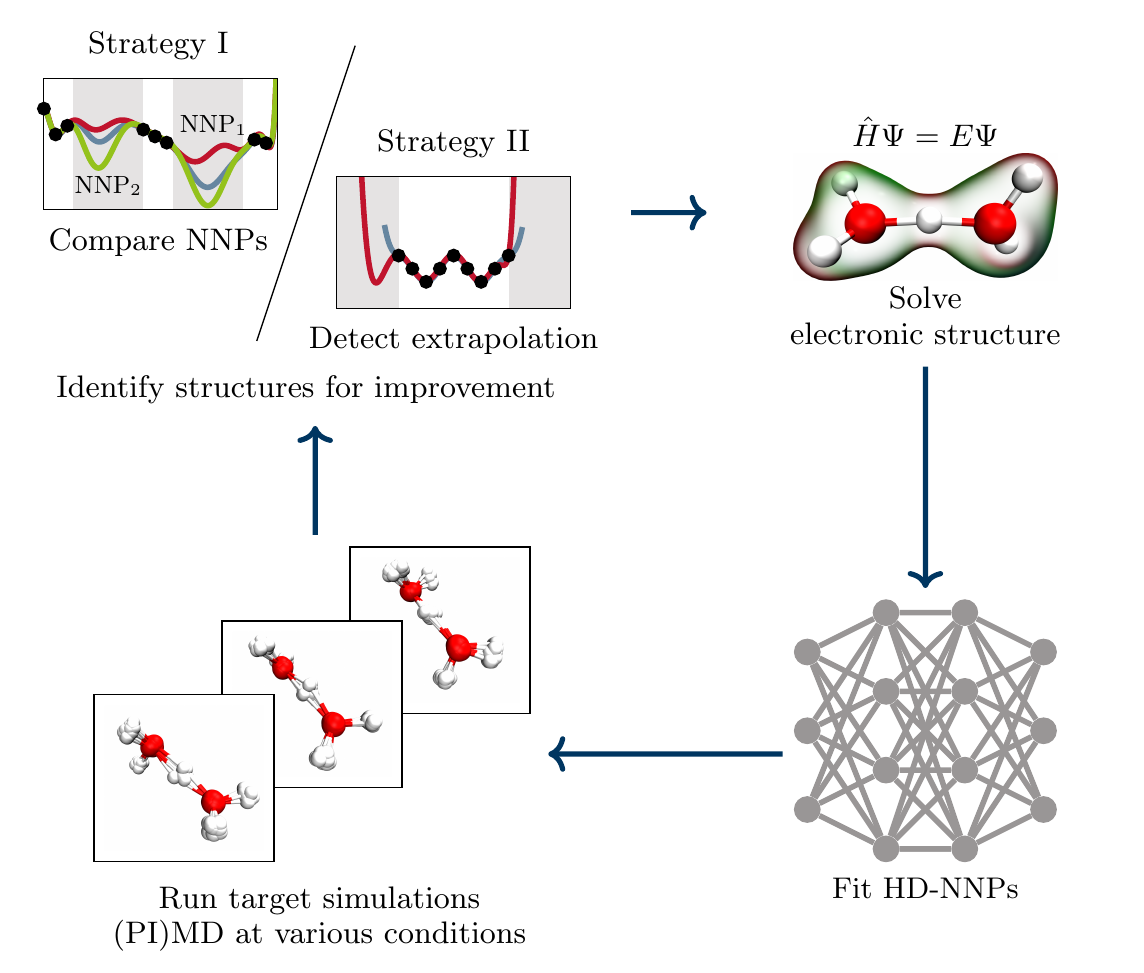}
    \caption{
        Schematic 
illustration
        of the four general steps in the developed automated
        fitting procedure. First, an ensemble of physically relevant structures 
        is generated
(lower left corner). 
        At the start of the automated process, this is achieved 
        with DFT based ``on--the--fly'' MD and PIMD simulation techniques, 
        while later the preliminary NNPs
        can be utilized for this step. Afterwards, relevant structures for an
        improvement of the potential energy surface are selected either by
        comparison of two NNPs 
(Strategy~I) 
        or by detection of points with extrapolation problems 
(Strategy~II) 
        to improve regions of the surface
        that are underrepresented in the training set,
highlighted in gray
(upper left corner). 
        Only for these selected 
        configurations, explicit electronic structure calculations with the
        chosen reference method, in this case coupled cluster theory,
        are performed 
(upper right corner)
        to provide additional reference values for the training set. 
        Afterwards, the 
two
NNPs are refitted to the enhanced
coupled cluster 
        training set 
(lower right corner) and the cycle starts over
by gathering
new physically relevant structures (lower left corner),
which can eventually be generated using the preliminary coupled cluster NNPs
once they are sufficiently stable to run MD and PIMD simulations.
    }
    \label{fig:afp}
\end{figure}

For the automated development of highly accurate potential energy
surfaces of finite sized clusters fitted to coupled cluster
reference calculations, we start by generating physically
meaningful structures by DFT, 
being an affordable, robust and general electronic structure approach, using both
\textit{ab initio} molecular dynamics (AIMD)
and \textit{ab initio} path integral MD (AI-PIMD)
on--the--fly simulations~\cite{Marx2009} 
to provide a set of 
very representative 
atomic configurations
for the specific system(s) of interest. 
Including PIMD sampling already at
this stage is important in order to provide
structures that feature the correct 
quantum fluctuations which are strongly
frequency dependent. 
In particular 
the zero-point vibrations and the associated energies
span a relatively large range 
in systems such as the present ones in view of the 
simultaneous presence of both, 
high-frequency small-amplitude modes 
(such as the intramolecular O--H stretches) 
and low-frequency large-amplitude modes 
(such as the intermolecular O$\cdots$O stretches).
As a result, such vastly different modes  
reach their classical limit at distinctly different temperatures
in PIMD simulations, 
whereas classical MD simply establishes plain 
energy equipartitioning at the given temperature.
Together, these classical and quantum
ensembles serve as the basis for an automated iterative
selection of the most relevant structures by repetition of the
following steps as schematically illustrated in Fig.~\ref{fig:afp}.
First, a 
quite
small number of 
structures is randomly extracted from these ensembles and
reference calculations 
using coupled cluster
theory are performed. 
Afterwards, two distinct NNPs are trained to these 
rather few
points.
These two potentials are used 
to predict the energies
of a large number of structures from the original 
DFT based
set of configurations, which is very inexpensive. 
In
a subsequent step, the training set
gets
further improved to account for differences 
with respect to the high level theory reference method.
To this end, only a 
few configurations with largest deviation between the
two predictions are selected (denoted as ``Strategy~I'' in Fig.~\ref{fig:afp})
to be included in the training set, thus computing
their (computationally demanding) coupled cluster energies. 
After having added these reference energies, the procedure starts
over again by fitting the next early generation of two NNPs. 
Importantly, these early generation NNPs need not to be optimized
to the level that they allow, by themselves, stable MD or PIMD simulations.

Using this approach, the most representative points of
the starting ensemble are selected in an unbiased and
efficient way.
In addition, this allows one to keep the number of
expensive coupled cluster reference calculations to a minimum.
However, the selected points are not yet optimal
for the high level reference method since DFT based
structures 
are expected to differ from those provided by the reference method, i.e. they are
biased with respect 
to those given by the desired (but unknown) 
coupled cluster surface.
Therefore, the 
boundaries 
of the network are improved by expanding the
configuration space of the training set
in a next step.
This is achieved by running a variety of
simulations at different conditions and with different sampling methods,
but now 
employing the preliminary 
coupled cluster based 
NNP. 
These simulations include all possible target approaches, such
as classical and path integral molecular dynamics and enhanced sampling
methods for which the final coupled cluster NNP is going to be used.
As before, PIMD sampling needs to be
included at this point if the quantum
nature of the nuclei shall be accounted for
in future uses of the NNP. 
This is necessary since classical MD 
can not provide the correct
frequency dependent quantum fluctuations
all the way from low (intermolecular) to high (intramolecular) frequencies, 
which usually leads to dissociation 
of molecular clusters 
if the temperature is increased 
up to the point to reach the classical limit
in case of the highest modes of the system. 

During these simulations, points that are subject to
extrapolation are identified by comparison of
the structural information encoded by the
atom--centered symmetry functions to the training set
(called ``Strategy~II'' in Fig.~\ref{fig:afp}).
In case the 
NNP based
simulation leaves the range of the
training set, that particular configuration is needed for an
improvement of the network and 
is selected 
for an explicit reference calculation.
After sufficiently many such points 
(but as few as e.g. 20) 
have been identified,
the simulations are aborted and
new coupled cluster reference calculations are performed to
improve the training set
only where required. 
Afterwards, a new NNP is fitted that now
has the capability to reliably predict the energy of
the selected configurations.
Thus, the simulations can be resumed to identify
further points needed for an improvement.
Utilizing the preliminary NNP for these simulations
provides several advantages.
First of all, a variety of different conditions
can be 
explored and representative configurations sampled
without problems due to the inexpensive
functional relation of the NNP that is many orders
of magnitude faster than the DFT based
description of the electronic structure.
This allows one, in addition, to reach rather long time scales
and to generate statistically uncorrelated new configurations.
At the same time, the preliminary NNP is approaching
the coupled cluster reference over time, which ensures
convergence.
In 
practice, the majority of reference
points is selected based on strategy~I, since
the DFT based structures usually provide a
rather balanced sampling of configuration space.
Still, strategy~II is fundamentally important
in order to gauge the NNP towards the coupled cluster PES
and to provide truly uncorrelated new structures
at various conditions.

After sufficiently many uncorrelated representative
structures at different conditions and with a variety of
sampling techniques have been generated, the
training set needs to be further refined as shown in the
following.
While the boundaries of the network were iteratively
improved, it is still possible that ``holes''
are present in the data set.
Therefore, the above described
automated comparison of two networks
is utilized again to identify regions in the training
set that are not yet optimally represented.
This is done for each of the different sampling
techniques and conditions individually for
the following reasons.
First of all, the separated improvement allows
for a high degree of parallelization
of the search for additional structures, which speeds
up the fitting procedure.
In addition, the different simulations may
sample various regions of the PES
with very distinct energies.
Thus, the separate improvement prevents that regions
with rather small energy differences remain undetected.
In this process, it is important to have sufficient overlap
of the different ensembles in order to prevent
``holes'' in the training set.
The automated process is completed once
the difference of the networks for the predicted energies
converges for all ensembles of structures.
This indicates that the most representative
configurations have been identified for all distinct
conditions.
Therefore, the different training points of all
ensembles can be combined for a final fit of the NNP.
\section{Computational details}
\label{sec:comp-det}

To generate the required initial set of reference structures
as a starting point for our automated fitting procedure
as described in the previous section, 
AIMD  and AI-PIMD 
simulations~\cite{Marx2009} of the protonated water clusters
were performed
from \cf{H3O+} to \cf{H9O4+}, including also
the uncharged \cf{H2O} monomer.
These simulations have been carried out with our in--house
developer's version of the 
\texttt{CP2k} program package~\cite{CP2K,Hutter2014/10.1002/wcms.1159} 
in 9 to 20~\AA{} cubic boxes with nonperiodic cluster boundary conditions.
The electronic structure
was described by the RPBE exchange correlation
density functional~\cite{Hammer1999/10.1103/PhysRevB.59.7413}
together with the D3 dispersion
correction~\cite{Grimme2010/10.1063/1.3382344} 
using the two--body terms 
and zero damping
as evaluated on--the--fly
using the \texttt{Quickstep} module~\cite{VandeVondele2005/10.1016/j.cpc.2004.12.014}.
The charge density was represented on a grid up to a plane wave cutoff of 500~Ry.
The TZV2P basis set together with Goedecker--Teter--Hutter
pseudopotentials to replace the
core electrons of the oxygen atoms~\cite{Goedecker1996/10.1103/PhysRevB.54.1703}
was used for the description of the Kohn--Sham orbitals.
The SCF cycles were converged to an error of $\epsilon_\text{SCF} = 10^{-7}~\text{Ha}$.
This electronic structure setup has been shown repeatedly to provide
reliable properties of water and is therefore the ideal
starting point for our fitting 
procedure~\cite{Morawietz2013/10.1021/jp401225b,
Forster-Tonigold2014/10.1063/1.4892400,
Morawietz2016/10.1073/pnas.1602375113}.
For each molecule at least 
\SI{100}{\pico\second} AIMD and
\SI{25}{\pico\second} AI-PIMD trajectories
were generated 
(after additional \SI{2.5}{\pico\second} equilibration periods in each case) 
to sample representative configurations
with a time step of \SI{0.25}{\femto\second}
at a temperature of \SI{300}{\kelvin},
where in the case of the AI-PIMD simulations the path integral
has been discretized using 6 Trotter replicas.
In case of the AI-PIMD simulations, the PIGLET
algorithm~\cite{Ceriotti2012/10.1103/PhysRevLett.109.100604}
was applied to sample the canonical quantum distribution,
while for AIMD a massive Nosé--Hoover chain thermostat with
a chain length of 5 was employed.
For \cf{H9O4+}, simulations starting from the four established
isomers were performed in order to generate relevant
configurations also for the higher energy isomers.
From these reference ensembles we extracted
configurations separated by \SI{10}{\femto\second} that 
serve as the basis for the generation of the NNP potentials
based on DFT 
as outlined above.

The accurate reference energies were calculated
via CCSD(T) by employing the explicitly correlated
F12a method~\cite{Adler2007/10.1063/1.2817618,
Knizia2009/10.1063/1.3054300}
to correct for the basis set incompleteness error.
In addition, size consistent scaling of the triples
suggested in Ref.~\citenum{Knizia2009/10.1063/1.3054300} was
employed together with the aug-cc-pVTZ basis
set~\cite{Dunning1992/10.1063/1.462569,
Woon1994/10.1063/1.466439}
for the calculation of the energies.
This so-called
CCSD(T*)-F12a/aug-cc-pVTZ
electronic structure setup 
has been shown to provide energies
close to the 
complete
basis set 
(CBS)
limit~\cite{Knizia2009/10.1063/1.3054300}.
All reference calculations were performed with the
\texttt{Molpro} program package\cite{MOLPRO}.

The NNP architecture is 
designed
as detailed in the following.
A set of symmetry functions for each element are
chosen to transform the coordinates of the system
to the input vectors for the atomic NNs.
We chose the parameters of the
symmetry functions for the final NNP
according to Ref.~\citenum{Morawietz2018/10.1021/acs.jpclett.8b00133},
which have been optimized for the description of water.
The
largest cutoff in this set is
about
6.5~\AA{} for both, oxygen
and hydrogen atoms.
Therefore, the set of symmetry functions provide
a global description of the
chosen clusters
since no atoms outside the cutoff radii are present,
which could give rise to long-range electrostatic 
interactions that could not be covered by the NNP.
The values of each symmetry function were centered
around the respective average value of the training set and normalized to
values between zero and one.
These vectors serve as the input for the atomic NNs,
which consist in all cases of two hidden layers with 30 nodes each,
and yield the atomic energy contributions
that sum up to the total energy.
Bias nodes with weight parameters $b$ were attached
to all layers but the input layer.
The hyperbolic tangent was employed in all hidden layers
except the output layer, where  a linear activation function
for the output layer prevents a confined range of output values.
The NNPs are constructed by first splitting the
set of 
CCSD(T*)-F12a/aug-cc-pVTZ
reference data
into a training set (90\%)
and an independent test set (10\%).
Subsequently, the weight parameters of the NNs were iteratively optimized to
minimize the error of the training set, while the test set provides an estimate 
for the transferability to structures not included
in the training set and is used to detect over fitting.
Learning was achieved by
optimizing the weights according to the
adaptive global extended Kalman filter~\cite{Shah1992/10.1016/S0893-6080(05)80139-X,
Blank1994/10.1002/cem.1180080605,Witkoskie2005/10.1021/ct049976i}
as implemented in our in--house program \texttt{RuNNer}~\cite{Behler/Runner}.
Further details including the complete description
of the NNP and all of its optimized parameters
can be found in the supporting information.

The MD and PIMD simulations at different conditions
using the generated NNPs were performed 
with our in-house  NNP extension of the
\texttt{CP2k} program package~\cite{CP2K,Hutter2014/10.1002/wcms.1159}.
These consist of simulations at temperatures of 10, 30, 70, 100, 200,
300, and \SI{600}{\kelvin} for the MD simulations
with otherwise the same settings as 
specified before for the AIMD simulations. 
For the PIMD simulations sampling of the
quantum partition function at 1.67, 50,
and \SI{300}{\kelvin} was performed
using PIGLET thermostatting with
512, 124, and 16 Trotter beads, respectively.
During these MD as well as PIMD simulations, 
extrapolated points were identified
when two consecutive structures
were outside the range of the symmetry functions in the training set
of the networks.
In such a case, the simulations were aborted and
new simulations were automatically started with a different
random number seed.
New electronic structure calculations were performed
after 20 
such extrapolating
structures have been selected and the NNPs
were re-fitted to the updated training set.
Finally, a run time up to \SI{8}{\nano\second}
with a time step of \SI{0.5}{\femto\second}
and \SI{0.5}{\nano\second} with a time step
of \SI{0.25}{\femto\second} under all above mentioned conditions
were generated without occurrence of extrapolations
for the classical and quantum description of
the nuclei, respectively, using NNP-MD and NNP-PIMD.

In the case of the tetramer, additional enhanced sampling
simulations based on constrained MD techniques were
performed to incorporate rearrangements
between the different isomers
which involve high energy barriers.
After careful analysis of the minimum
energy pathway, obtained by the zero
temperature string method~\cite{Ren2002/10.1103/PhysRevB.66.052301},
the OOO angle $\angle_\text{OOO}$ was identified as the
most promising reaction coordinate
to describe the isomerization
of the Eigen complex to the Zundel--like state.
Classical MD and quantum PIMD simulations
were constrained via the ``RATTLE''
algorithm~\cite{Andersen1983/10.1016/0021-9991(83)90014-1}
from 30 to 125$^\circ$ in 40 angular steps.
MD simulations were performed at temperatures
of 200, 100, and \SI{50}{\kelvin} via thermostatting
using a Nosé--Hover chain thermostat.
In the case of the PIMD simulations,
the constraint was applied exclusively to the centroid
following earlier work on free energy calculations
in the path integral formalism~\cite{Gillan1987/10.1088/0022-3719/20/24/005,
Voth1989/10.1063/1.457242,
Walker2010/10.1063/1.3505038}.
Quantum simulations at 200 and \SI{100}{\kelvin} were
performed by thermostatting via the PILE
Langevin thermostat~\cite{Ceriotti2010/10.1063/1.3489925},
were the path integral was discretized using 
32 and 64 replicas, respectively.

In order to validate the accuracy of
the final NNP for production runs,
classical MD simulations
were performed at \SI{300}{}
and \SI{600}{\kelvin}, while
quantum PIMD simulations were carried
out at \SI{1.67}{} and \SI{300}{\kelvin}.
All
trajectories were propagated for
\SI{25}{\pico\second} with otherwise
the same settings as for
the above mentioned simulations.
Afterwards, the last \num{100} points
of these trajectories were re-evaluated
with the coupled cluster reference
method to directly compare
the performance of the NNP to the exact reference.

\section{Neural Network Potential of Protonated Water Clusters}
\label{sec:res}
\subsection{Automated Neural Network Fitting Procedure}
\label{ssec:fit}

In order to develop a neural network potential that can be applied
for the description of protonated water clusters from the monomer, \cf{H3O+},
up to the tetramer, \cf{H9O4+}, DFT based
simulations were performed at \SI{300}{\kelvin} to sample both the classical
and quantum configuration space of all clusters
using on--the--fly AIMD and AI-PIMD simulations, respectively. 
The water monomer, \cf{H2O}, was also explicitly considered in these
simulations to provide the correct description of the
dissociation products of the protonated water clusters
in the final NNP. 
For the largest cluster, \cf{H9O4+}, all four known isomers (Eigen, Ring,
Zundel-c and Zundel-t, see 
below for details) 
were used as the starting point of these simulations.
Afterwards, 
at least
\num{100000} uncorrelated
structures 
are
extracted from these simulations
for each cluster to
serve as the basis of the automated NNP fitting procedure
as outlined in the previous section.
For every cluster size, the iterative selection of the most relevant
configurations both for the classical and the quantum DFT
ensembles were performed using at least \num{80} refining stages.
In each refining stage, the 
energies
of \num{50000}
configurations
were
evaluated with two NNPs and
the \num{20} points with largest energy differences
were added to the training set except at the beginning,
when
\num{20} random structures were chosen to start.
Afterwards, the training sets of the classical and quantum ensembles
were combined and preliminary
neural networks were fitted for 
each
cluster individually.
In the next step, the 
boundaries 
of these networks were 
expanded
in order to
extend
the configuration space of the reference set
by running classical and quantum trajectories at different conditions.
These consist of simulations at temperatures of 10, 30, 70, 100, 200,
300, and \SI{600}{\kelvin} for classical nuclei, as well as
1.67, 50, and \SI{300}{\kelvin} for quantum nuclei.
During these simulations, extrapolated points were identified
and iteratively added until
sufficiently long trajectories
under all above mentioned conditions
were generated for the classical and quantum description of
the nuclei, respectively.
Recall
that only a few configurations
are selected based on this strategy, while
the major contribution of reference points
is selected based on the comparison of two NNPs.
However, it is still a fundamentally important
step in order to gauge the DFT based structures
to the coupled cluster configuration space.
In the case of the tetramer, additional enhanced sampling
simulations based on thermodynamic integration were
performed to incorporate rearrangements
via higher--lying energetic barriers 
between the different isomers.
All the computational details of these simulations
and also the neural network specification as well
as details on the coupled cluster reference calculations
are presented in Sec.~\ref{sec:comp-det}.
In a next step, the training sets of the respective clusters
were refined  by the same iterative comparison of two networks
as before until the differences between the networks was
converged.
Afterwards, all reference calculations of all clusters
were combined for a final fit of the neural network potential.
In order to investigate if this final NNP 
has reached its asymptotic limit
with respect to the size of the training set,
or if the model would benefit from additional
data, we analyzed the learning curves
for the final data set in the supporting
information.
This analysis shows that the model
is very close to its asymptotic accuracy
and the automated fitting protocol is
therefore considered to be converged.

\begin{table}[t]
  \caption{
        Root mean square error
        of the energies per atom for the
        full data set (``All'') and for each individual cluster size.
        The values refer to the training set, while the numbers in
        parentheses correspond to the structures in the independent
        test set which are not 
        considered in the fit of
        the neural network
        and thus provide an unbiased estimate of the
        predictive power.
  }
  \label{tab:qual}
  \begin{tabular*}{0.9\linewidth}{@{\extracolsep{\fill}}ccc}
    \hline
    Data & Number        & RMSE        \\
    set  & of structures & (kJ/mol)/atom \\
        \hline
      \centering
    All         &    49242  (5470)  &   0.06    (0.08)  \\
 \noalign{\smallskip}
    \cf{H2O}    &   5933    (660)   &   0.03    (0.03)  \\
    \cf{H3O+}   &   6489    (711)   &   0.05    (0.05)  \\
    \cf{H5O2+}  &   8869    (1057)  &   0.07    (0.09)  \\
    \cf{H7O3+}  &   9356    (973)   &   0.07    (0.08)  \\
    \cf{H9O4+}  &  18595    (2069)  &   0.06    (0.10)  \\
    \hline
  \end{tabular*}
\end{table}

Overall, around \num{55000} configurations 
for which coupled cluster energy calculations were carried out have been
selected during the automated fitting procedure for all clusters
in total.
Roughly \num{50000} of these points
were used for the training of the final network, while
the remaining \num{5000} configurations
have been
used to
access the transferability of the fit in an independent
test set.
Compared to standard approaches for the representation
of a comparable PES,  
considerably less computationally demanding 
reference calculations were required to converge
the PES, which is a considerable reduction of the computational cost.
This clearly highlights the advantages of
the automated selection of the training points
over traditional approaches based for example on grids.
The contribution of each individual cluster to the total
training and test set is summarized in
Table~\ref{tab:qual}.
As seen therein, the higher number of degrees
of freedom with increasing cluster size
is well incorporated in the final reference set.
In addition, the root mean squared error (RMSE)
per atom is included
in the table which for the total fit is around
\SI{0.06}{\kilo\joule\per\mol} per atom.
This error corresponds to a total energy error of
\SI{0.8}{\kilo\joule\per\mol} and is thus 
much better than chemical accuracy (being 1~kcal/mol or about \SI{4}{\kilo\joule\per\mol}) 
even for the largest considered cluster,
which thus underlines the high quality of the fit.
Furthermore, the fitting error is stable
over the different cluster sizes, indicating that
all of them are represented with similar accuracy. 
Last but not least, the
error does not deteriorate significantly
for the test set, which has not been considered
during the fit, and thus provides an estimate
of the transferability of the fitted potential.

\begin{figure}[t]
    \centering{}
    \includegraphics[width=1.0\linewidth]{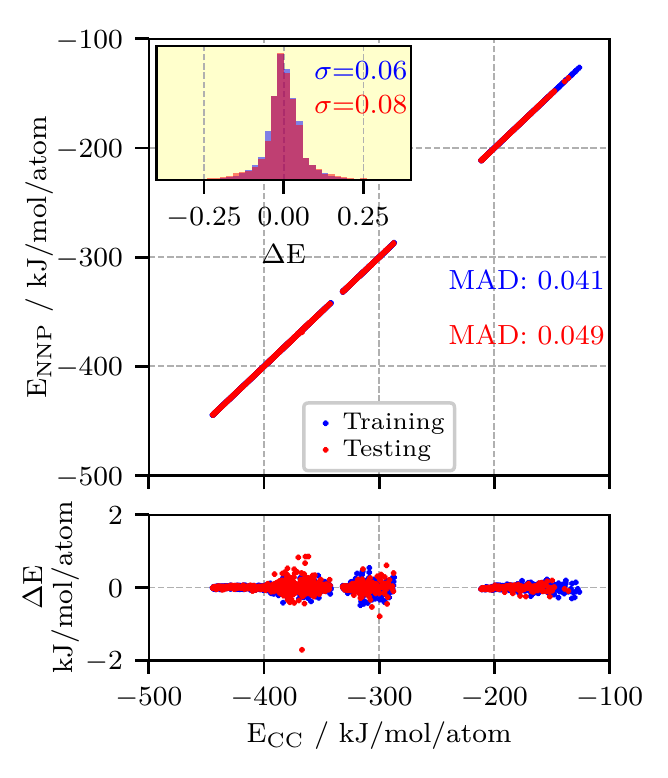}
    \caption{
        Correlation of the energy per atom from explicit
        CCSD(T*)-F12a/aug-cc-pVTZ
        calculations (CC) and the NNP predictions
        for the final reference data set, see text.
        The mean absolute difference (MAD) for the training and test
        set are given in blue and red, respectively.
        The lower panel shows the energy differences between
        coupled cluster reference and NNP prediction over
        the whole range of reference energies,
        while the inset 
in the upper panel
        shows the histograms 
of 
        the energy differences
        including the corresponding standard deviations $\sigma$ in
        the respective color. 
    }
    \label{fig:prot_fit}
\end{figure}
The resulting correlation between the reference
CCSD(T*)-F12a/aug-cc-pVTZ
(see Sec.~\ref{sec:comp-det} for details) energies per atom and the final NNP
energies
for the training and test sets is shown in Fig.~\ref{fig:prot_fit}.
All considered points feature almost perfect correlation over the
full range of energy values.
Note that the gaps in the range of energies are a result of
describing different cluster sizes with one and the same NNP,
which is thus 
demonstrated to be 
capable of describing these very distinct
clusters with one set of 
fitted
parameters.
The lower panel of the figure 
quantifies the 
deviation of the predicted energies over the whole range
of the reference set.
These are consistently small, for both the training and test set
shown in blue and red colors, respectively. 
To further illustrate the accuracy of the fit,
the histogram 
of
these energy differences is shown
in the upper inset of Fig.~\ref{fig:prot_fit}.
It features a very narrow distribution for
both, the training and test set with a standard
deviation (which is equal to the RMSE)
of around \SI{0.08}{\kilo\joule\per\mol} per atom.
This error compares well with usual fitting errors
of similarly complex hydrogen bonded clusters.

Overall, these results demonstrate that NNPs 
obtained from an automated procedure
are able to represent highly complex potential energy landscapes
including different system sizes with very high precision
and are therefore ideally suited to
reproduce the coupled cluster reference landscape. 
At the same time, the evaluation of the network
is many orders of magnitude faster than the explicit
electronic structure calculation.
A reference calculation for the largest considered cluster,
\cf{H9O4+}, takes on the order of seven hours on a single core
with rather high memory demands.
The evaluation of the NNP can be performed in around
\SI{0.1}{\milli\second}, thus about $10^{8}$ times faster at roughly the
same accuracy.
This opens the door for systematic
investigations of these systems using essentially exact interactions
as provided by
CCSD(T*)-F12a/aug-cc-pVTZ
theory. 
The applicability of NNPs for the
description of differently sized clusters
has already been confirmed for DFT based potentials
of water clusters~\cite{Morawietz2012/10.1063/1.3682557,
Morawietz2013/10.1021/jp401225b} and also
for 
protonated water
clusters~\cite{Natarajan2015/10.1039/C4CP04751F}.
However, these networks were additionally optimized
to structure dependent forces acting on the nuclei
during the fit, producing
very similar results for the energies as presented above.
For the present study, the usage of forces
is unfortunately out of scope, since
the calculation of forces for coupled cluster theory are increasingly more
demanding especially for the largest considered cluster
which contains 13 atoms.
To specifically test, if forces would further improve the accuracy
of our fit, we therefore resorted to the following strategy:
For the hydronium cation, \cf{H3O+}, the evaluation
of the forces is still feasible using the
CCSD(T*)-F12a/aug-cc-pVTZ
method.
Therefore, the forces for all hydronium structures in
our data set were explicitly calculated
and two test networks were afterwards trained to this data.
The first one utilizes the forces during the optimization,
while the second one is trained to energies only.
Afterwards, the test and training error for
both fits can be compared to analyze the differences.
For both fits very similar test and training errors were obtained
for the energies.
In addition, the error on the forces
is very similar in both fits, although
the force information was not used for one of the networks.
The usage of forces, if available, would clearly
be able to further reduce the number of required reference
calculations
(yet at the expense of strongly increasing the computational effort
per reference configuration). 
However, they do not further improve
the quality of our fit, if already sufficiently many
training points are in the reference set.

In summary, the extension to the coupled cluster reference
using
only
the training on energies 
provides results almost
indistinguishable from the reference method.
In addition, the automated assembly of the training set
ensures that the procedure selects sufficiently
many points for the representation of the PES
and significantly reduces the number of required reference
calculations.
It also allows one to easily transfer this methodology
to other systems and to readily develop new potentials. 

\begin{table}[p]
  \centering
  \caption{
      Binding energy $E_\text{bind}$ in kcal/mol
      (units were chosen for direct
      comparison with Ref.~\citenum{Heindel2018/10.1021/acs.jctc.8b00598})
      for selected stationary--point structures of
      protonated water clusters from \cf{H3O+} to \cf{H9O4+}
      optimized using the neural network potential (NNP)
      and the
      CCSD(T*)-F12a/aug-cc-pVTZ
      reference (CC).
Here, 
      $|$NNP-CC$|$ is the absolute difference between the
      two methods
and 
      $M$ is the number of water monomers in the cluster.
      Note that for the hydronium cation ($M=1$), 
$E_\text{bind}$ denotes the energy difference between \cf{H3O+}
      and the water monomer.
      The structures optimized with the NNP
      are depicted (those from CC optimization would
      be indistinguishable), were oxygen atoms are red and
      hydrogen atoms are white.
      The labels that are used throughout the text
      include an asterisk, if the shown
      configuration is the global minimum of the
      respective cluster size.
      }
    \label{tbl:stat_points}
    \begin{tabular}{c c c c r}
        \hline
        \noalign{\smallskip}
        $M$ & Structure & Label & Method & $E_\text{bind}$\\
        \noalign{\smallskip}
        \hline
        \noalign{\smallskip}
        \centering
        1 & \multirow{3}{3cm}{\centering{}\includegraphics[width=0.05\textwidth]{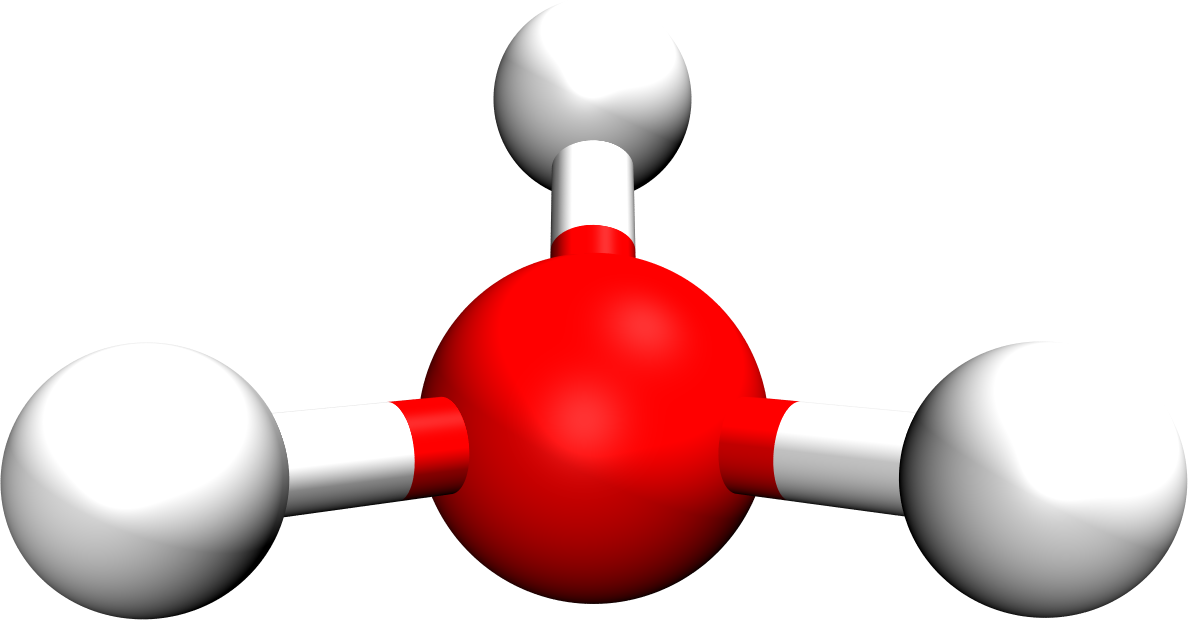}}
          &                   & NNP              & -171.4 \\
        & & C$_\text{3v}^{*}$ & CC               & -171.5 \\
        & &                   & $|$NNP-CC$|$     &    0.037 \\
        \noalign{\smallskip}
        \hline
        \noalign{\smallskip}
        \centering
        2 & \multirow{3}{3cm}{\centering{}\includegraphics[width=0.09\textwidth]{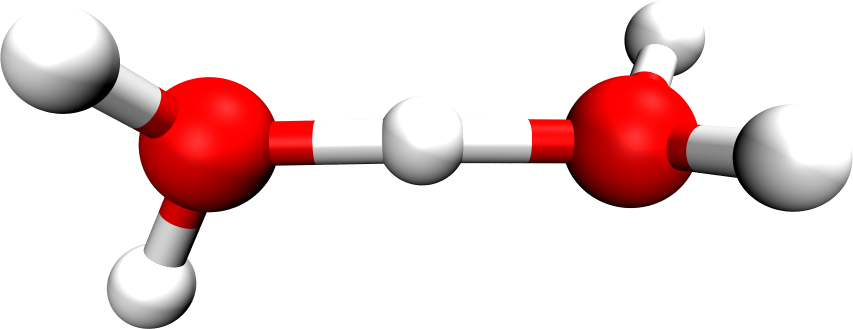}}
          &                   & NNP              & -34.0 \\
        & & C$_\text{2}^{*}$  & CC               & -34.0 \\
        & &                   & $|$NNP-CC$|$     &   0.084 \\
        \noalign{\smallskip}
        \hline
        \noalign{\medskip}
        3&\multirow{2}{3cm}{\centering{}\includegraphics[width=0.12\textwidth,angle=2]{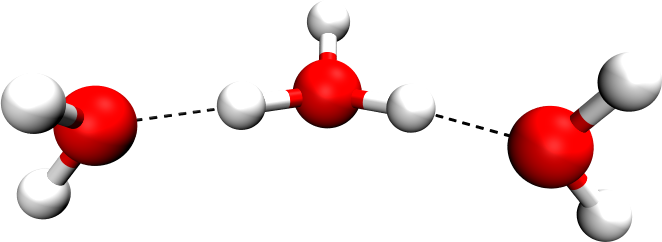}}
          &                   & NNP              &  -57.7\\
        & & W3$^{*}_{1}$      & CC               &  -57.6\\
        & &                   & $|$NNP-CC$|$     &    0.044\\
        \noalign{\smallskip}
        \hline
        \noalign{\medskip}
        3&\multirow{2}{3cm}{\centering{}\includegraphics[width=0.12\textwidth,angle=0.5]{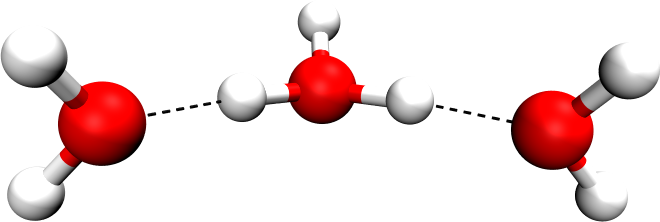}} 
          &                   & NNP              & -57.7 \\
        & &      W3$_{2}$     & CC               & -57.6 \\
        & &                   & $|$NNP-CC$|$     &   0.063 \\
        \noalign{\smallskip}
        \hline
        \noalign{\medskip}
        4&\multirow{2}{3cm}{\centering{}\includegraphics[width=0.12\textwidth]{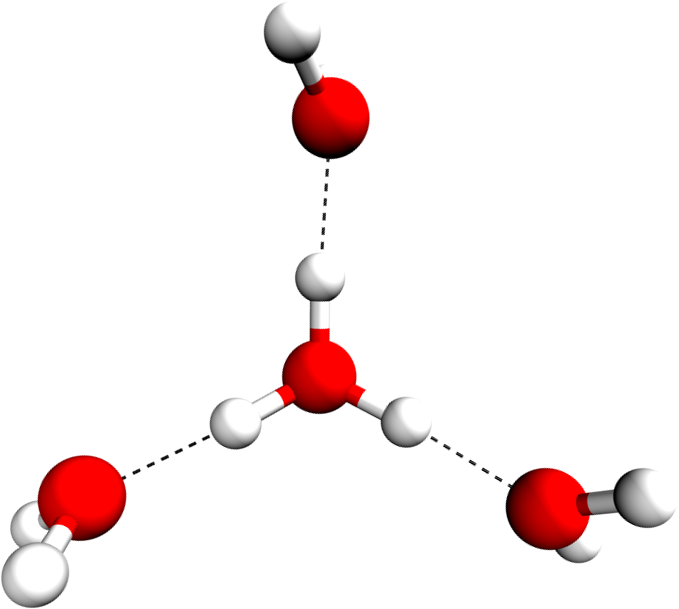}} &
                        & NNP              & -77.5 \\
        & & Eigen$^{*}$ & CC               & -77.5 \\
        & &             & $|$NNP-CC$|$     &   0.017 \\
        \noalign{\bigskip}
        \noalign{\bigskip}
        \noalign{\bigskip}
        \hline
        \noalign{\medskip}
        4&\multirow{2}{3cm}{\centering{}\includegraphics[width=0.12\textwidth]{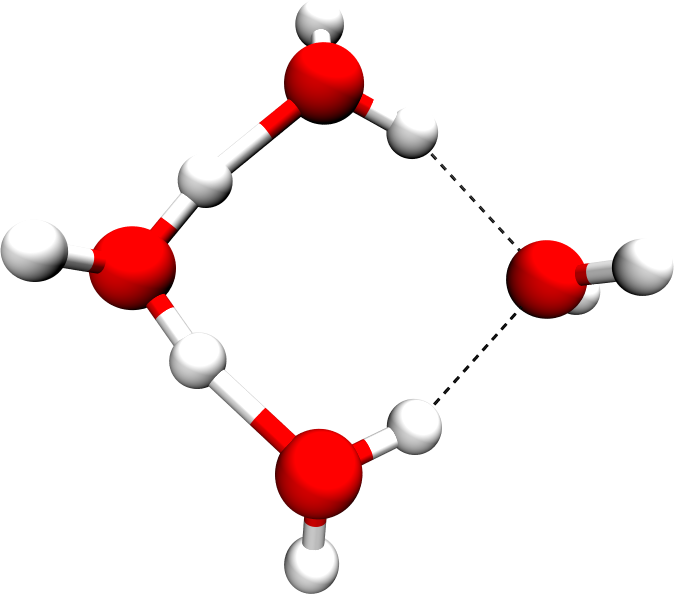}} & 
                        & NNP              & -73.5 \\
        & &Ring         & CC               & -73.5 \\
        & &             & $|$NNP-CC$|$     &   0.021 \\
        \noalign{\bigskip}
        \noalign{\bigskip}
        \noalign{\medskip}
        \hline
        \noalign{\medskip}
        \noalign{\medskip}
        4&\multirow{2}{3cm}{\centering{}\includegraphics[width=0.13\textwidth]{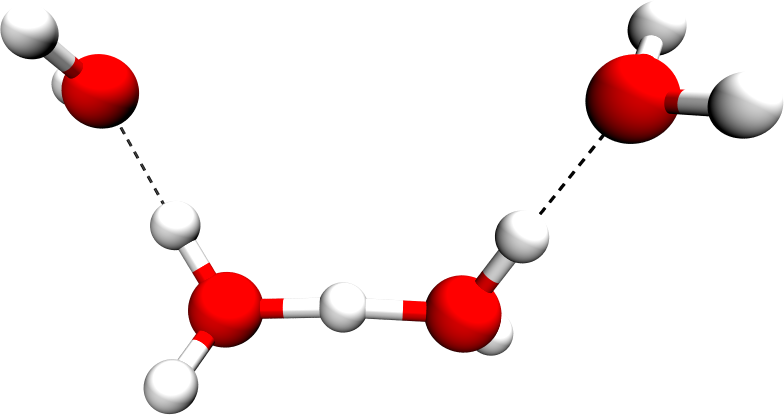}} & 
                        & NNP              & -73.6 \\
        & &Zundel-c     & CC               & -73.6 \\
        & &             & $|$NNP-CC$|$     &   0.014\\
        \noalign{\bigskip}
        \noalign{\medskip}
        \hline
        \noalign{\medskip}
        \noalign{\medskip}
        4&\multirow{2}{3cm}{\centering{}\includegraphics[width=0.13\textwidth]{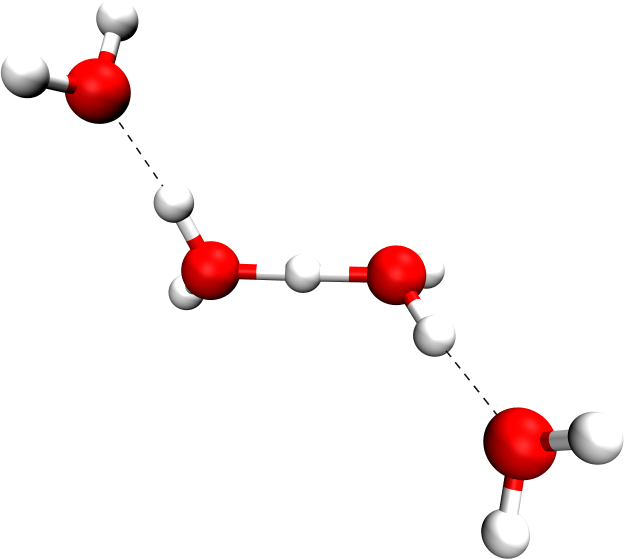}} & 
                        & NNP              & -73.5\\
        &&Zundel-t      & CC               & -73.5 \\
        & &             & $|$NNP-CC$|$     &   0.061 \\
        \noalign{\bigskip}
        \noalign{\bigskip}
        \noalign{\bigskip}
        \hline
    \end{tabular}
\end{table}

\subsection{Stationary Points and Vibrational Frequencies}
\label{ssec:sp_fr}
To further 
validate
the quality of the final neural network PES,
structural and energetic 
properties of a set of optimized stationary--point structures 
obtained from the NNP are compared to the
CCSD(T*)-F12a/aug-cc-pVTZ
reference data (CC).
In Table~\ref{tbl:stat_points} the binding energies of
the local minima for the hydronium cation, \cf{H3O+}, up
to the protonated water tetramer, \cf{H9O4+},
are provided for both methods.
For that purpose, all shown stationary points
were first optimized with the NNP and
then reoptimized using the coupled cluster reference,
resulting into essentially indistinguishable structures for all species.  
The binding energy, calculated as
usual from 
\begin{align}
    E_\text{bind}(\text{H}^{+}(\cf{H2O})_{n}) =& E(\text{H}^{+}(\cf{H2O})_{n}) -\nonumber \\
    &(n-1)E(\cf{H2O}) - E(\cf{H3O+})
\end{align}
based on the energies of the individually optimized clusters
as obtained using the respective methods,
is in very good agreement between the coupled cluster
reference and the NNP.
Even the
largest energy difference between
CCSD(T*)-F12a/aug-cc-pVTZ
reference
and NNP is found to not exceed \SI{0.1}{\kilo\cal\per\mol}. 
This fitting error is already 
in
the order of the remaining intrinsic uncertainty
of the F12~coupled cluster method that is used
and is at the level of state--of--the--art PES fits
such as those relying on
permutationally invariant polynomials~\cite{Braams2009/10.1080/01442350903234923}.
Last but not least,
the remaining fitting error does not
systematically 
deteriorate 
with increasing
system size, indicating very high quality representations 
for all included cluster sizes and isomers within a single NNP.

\begin{figure}[t]
    \centering{}
    \includegraphics[width=1.0\linewidth]{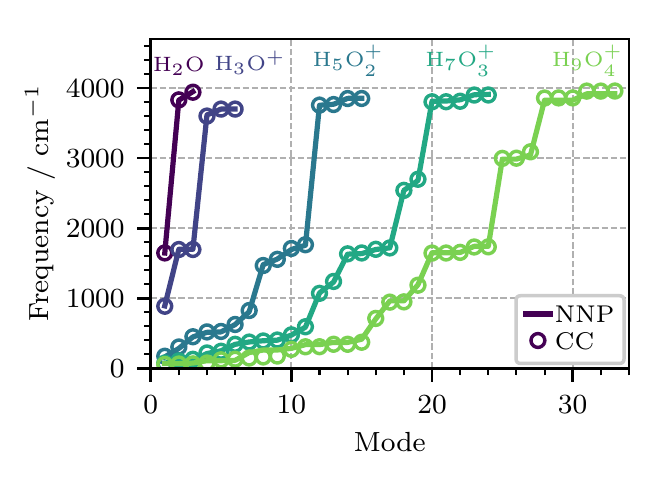}
    \caption{
        Comparison of the harmonic normal mode frequencies of the
        equilibrium structures corresponding to the global minima (depicted in Fig.~\ref{tbl:stat_points}) as obtained using
        the neural network potential (NNP) and the coupled cluster (CC)
        reference.
        All global minima have been optimized with the respective
        method.
    }
    \label{fig:nm_comp}
\end{figure}

Since the NNP has been optimized using only
the energy, the first derivative, i.e.
the forces, are also important quantities to be tested.
This particular validation has been done explicitly for the
hydronium cation, but is out of scope for the larger clusters.
Let us therefore focus on the curvature of the PES
around the global minima of the different clusters,
which is directly probed by the harmonic
normal mode frequencies.
As can be seen in Fig.~\ref{fig:nm_comp},
all investigated clusters show almost perfect
agreement between the coupled cluster reference
and the NNP.
Overall, the mean absolute difference between
the two methods is around \SI{10}{\per\centi\metre}, 
which furthermore supports the rather high quality of the fit;
note that this corresponds 
to about \SI{0.1}{\kilo\joule\per\mol}
or \SI{0.03}{\kilo\cal\per\mol} and is thus in line with
the reported energy errors.
This is a strong indication that the forces
of all clusters are also well described by the
final NNP, which is in agreement with previous
studies on a wide range of systems showing usually
a very good representation of the forces.

\subsection{Potential Energy Scans}
\label{ssec:pes_scan}
\begin{figure}[t]
    \centering{}
    \includegraphics[width=1.0\linewidth]{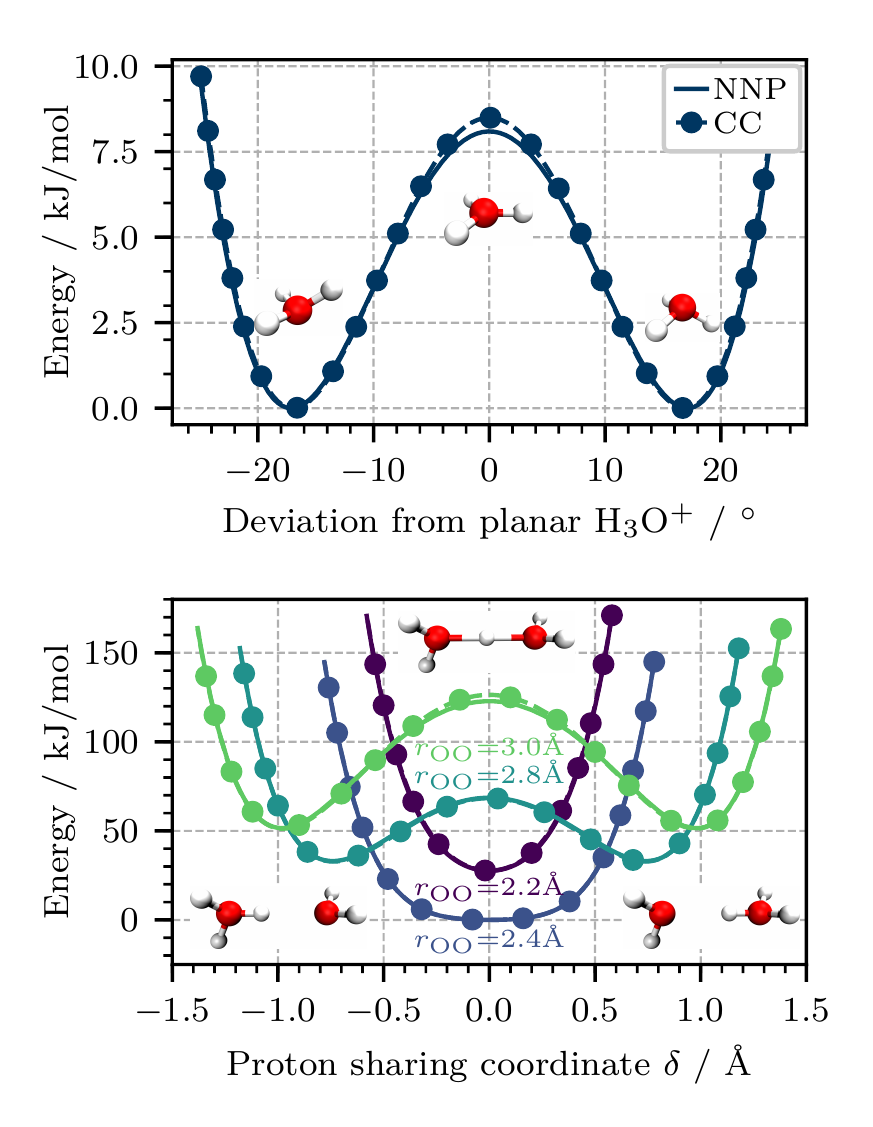}
    \caption{
        Scans of the potential energy profile 
obtained from the
        neural network potential (NNP) 
in comparison to the
        coupled cluster (CC) reference 
(marked with circles)
        for the hydronium and Zundel cation,
        \cf{H3O+} and \cf{H5O2+}, respectively.
        Top: Scan of the potential energy of the hydronium cation
        along the inversion pathway of the umbrella motion
        where the reaction coordinate is defined as the
        dihedral angle of the four atoms which provides
        the deviation from planarity of the molecule;
        note that the OH bond distances are kept at the value of the
        minimum energy configuration.
        Bottom: Scans of the potential energy of the Zundel cation
        along the proton sharing (transfer) coordinate $\delta$
        with the oxygen distance $r_\text{OO}$ constrained at different
        values
as indicated;
        note that the OH bond distances of the dangling hydrogen
        atoms and their orientation are conserved with respect
        to the minimum energy structure.
        All energies are shown relative to the respective 
        minimum energy
        structures
(which is given by the global minimum
around
$r_\text{OO}=2.4$~{\AA} for the Zundel cation).
    }
    \label{fig:pes_scan}
\end{figure}
\begin{figure}[t]
    \centering{}
    \includegraphics[width=1.0\linewidth]{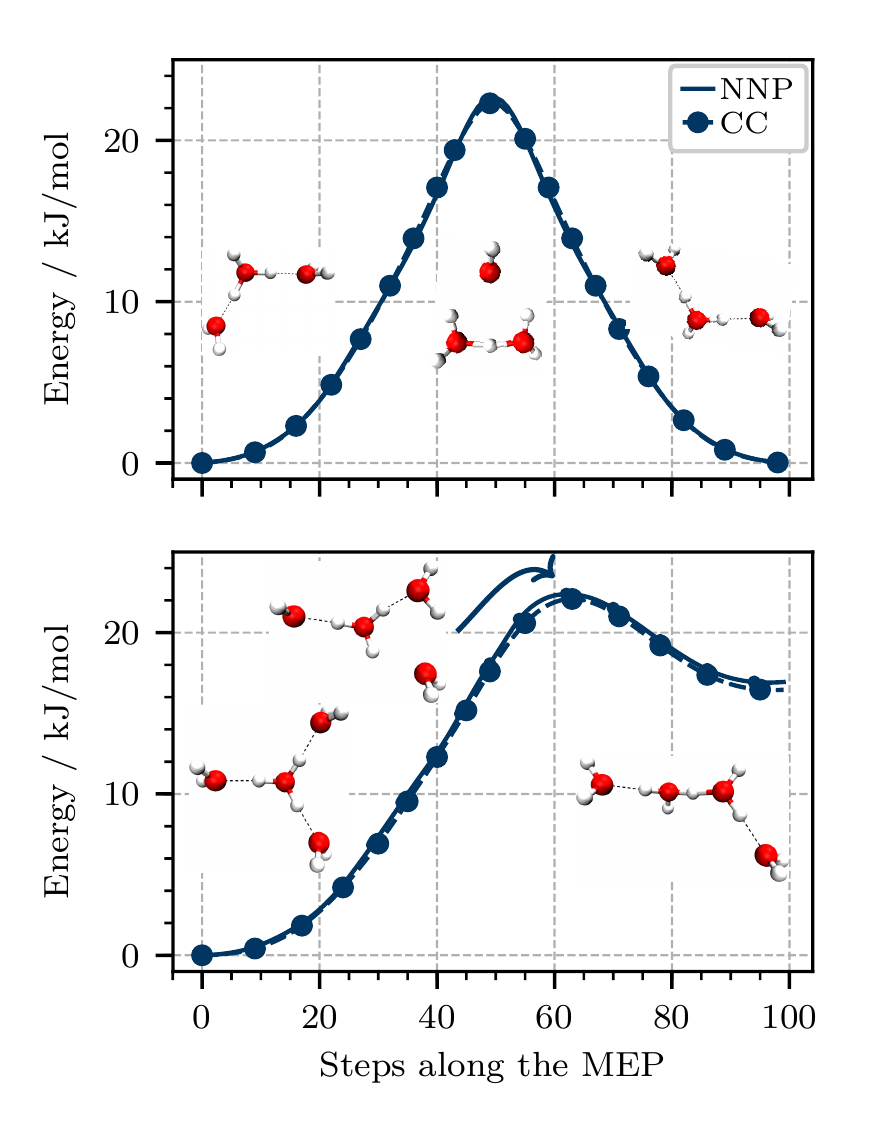}
    \caption{
        Potential energy profiles of the \cf{H7O3+} (top) and \cf{H9O4+} (bottom) 
        clusters along selected minimum energy paths (MEP)
for important isomerization reactions (represented by the respective reactant, transition
and product state configurations) 
        obtained using the zero temperature string method on the
        neural network potential (NNP).
        The coupled cluster (CC) reference was obtained by recomputing the
        energies along the NNP paths
at representative points and are marked with circles.
        The arrow in the bottom panel indicates the position
        of the upper-left configuration at the maximum of that reaction coordinate.
        All energies are shown relative to the respective equilibrium
        structures.
    }
    \label{fig:pes_zts}
\end{figure}
After having confirmed that stationary point structures and 
corresponding
harmonic frequencies are
well reproduced by the NNP, larger non--equilibrium
regions of the PES are now investigated in detail
and compared
to coupled cluster data. 
For that purpose, scans of the potential energy
along different reaction coordinates are
shown
for the hydronium and Zundel cation in Fig.~\ref{fig:pes_scan}.
For the hydronium complex, the 
umbrella-type
interconversion motion 
along the inversion coordinate was chosen,
which
results in a planar transition state with D$_\text{3h}$ symmetry.
As seen in the top panel of Fig.~\ref{fig:pes_scan}, the
energy profile as a function of the deviation from this planar
transition state results in two equal minima at around
$\pm 18^{\circ}$ that are separated by a barrier
of around \SI{8}{\kilo\joule\per\mol}.
The NNP is able to reproduce this profile with very good
accuracy, although small deviations 
from the coupled cluster reference
of about
\SI{0.3}{\kilo\joule\per\mol} 
can be detected in the transition state region,
yet these differences are well within chemical accuracy. 
\begin{figure*}[!t]
    \centering{}
    \includegraphics[width=1.0\textwidth]{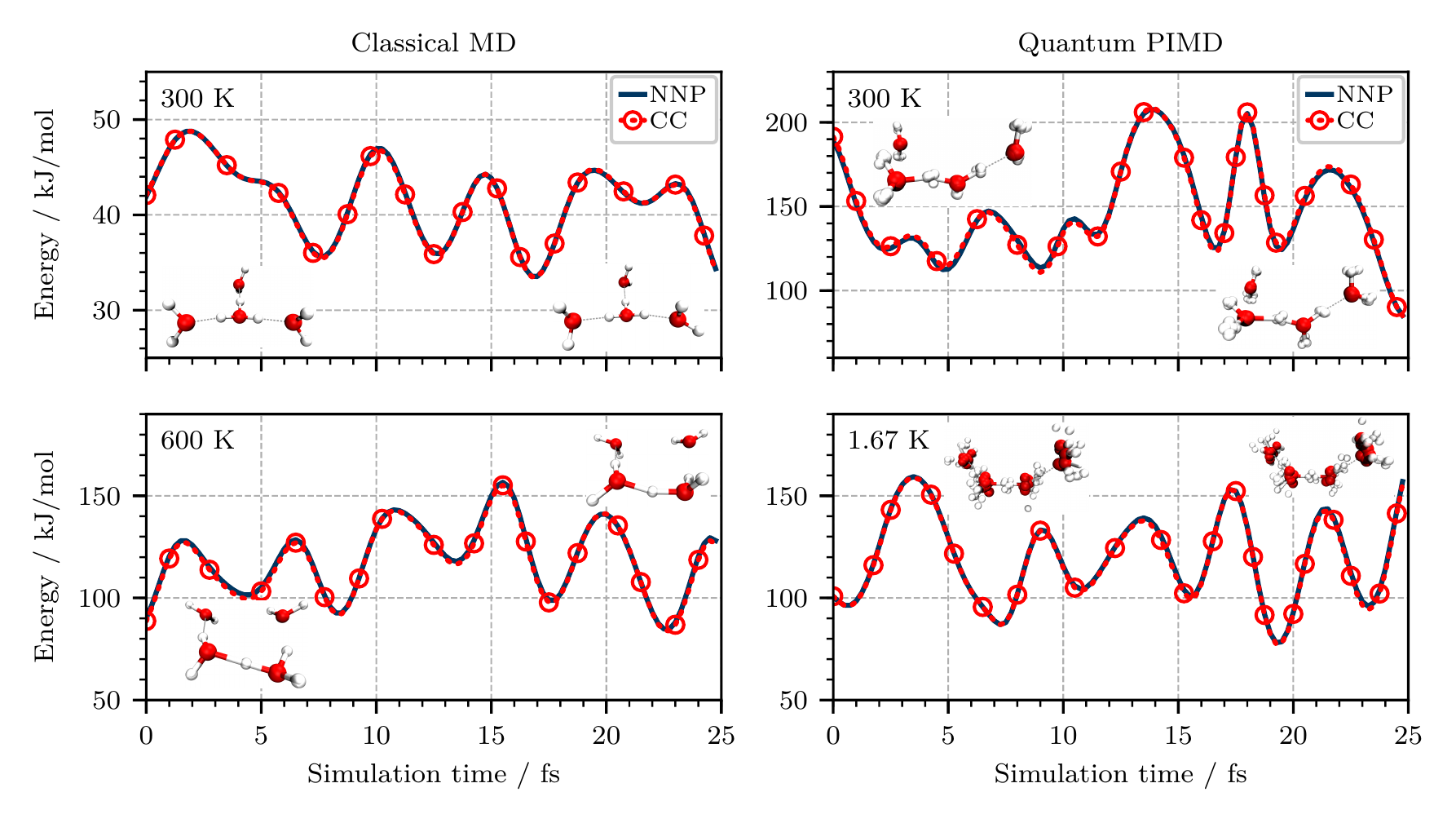}
    \caption{
        Potential energy along 
        classical MD (left)
        and 
one replica of
        quantum PIMD (right) trajectories
        of the
        protonated water tetramer at \SI{300}{\kelvin}
        (top) and \SI{600}{\kelvin} (bottom-left) as well as
        \SI{1.67}{\kelvin} (bottom-right)
in the Eigen (top-left), Zundel-c (right), and Ring (bottom-left) conformation 
        using the neural network potential (NNP).
        The coupled cluster (CC) reference was obtained by recomputing the
        energies along the NNP trajectories
        and are shown as red dotted lines (with a few circles
        added since the CC energies mostly superimpose the NNP data).
        All energies are reported relative to the Eigen equilibrium structure.
        The included snapshots depict the
initial and final configurations
        in the short trajectory window that is covered (see text). 
    }
    \label{fig:md_test}
\end{figure*}

For the Zundel cation, the 
energy profile along the 
proton transfer coordinate
$\delta$ (defined as the difference between the
two oxygen--proton distances) is 
computed for
different fixed oxygen--oxygen distances
as indicated in the figure.
By variation of the oxygen--oxygen distance it is
possible to drive this system from a symmetric 
single-well
hydrogen bond situation to an asymmetric
double-well energy profile 
for larger oxygen--oxygen distances
as seen in the bottom panel of Fig.~\ref{fig:pes_scan}.
The equilibrium 
structure
of this system 
corresponds to an OO~distance close to 2.4~\AA{} for which the 
broad anharmonic minimum with a centered proton position is observed in that figure. 
All other OO distances probe regions
higher in energy,
reaching relative energies of up to \SI{150}{\kilo\joule\per\mol}
and mapping a proton transfer barrier as high as \SI{75}{\kilo\joule\per\mol}
for the largest distance considered, $r_\text{OO}=3.0$~{\AA}.
For this system, essentially perfect agreement
between the NNP and the coupled cluster
reference is observed over
the
full energy range
explored. 
These different scans of potential energy profiles 
corresponding to large-amplitude rearrangements
show that
the NNP is able to describe very different
regions of the PES with convincing
accuracy for two representative species, \cf{H3O+} and \cf{H5O2+}.

In order to 
analyze
similar non--equilibrium regions also 
for the larger clusters, \cf{H7O3+} and \cf{H9O4+},
the zero temperature string method
was applied using the NNP to compute minimum energy paths 
(MEPs)
that connect two isomers of the different clusters.
With this approach, the
MEP
for the rearrangement of the protonated water trimer from the W3$_{1}$ isomer
to an equivalent W3$_{1}$ isomer with a different
central \cf{H3O+} unit 
(see top panel of Fig.~\ref{fig:pes_zts} for configurations) 
was obtained using the NNP.
For the protonated water tetramer,
a similar isomerization reaction transforms the
Eigen cation to a linear hydrogen bonded complex that contains a central 
Zundel-like motif as depicted in the bottom panel of Fig.~\ref{fig:pes_zts}.
Having determined the configurations along these MEPs based on the NNP,
a set of single-point
energies along these pathways were calculated using the coupled
cluster reference method. 
The resulting NNP {\em versus} CC comparison of the two potential
energy profiles is depicted in Fig.~\ref{fig:pes_zts}.
In case of
the protonated water trimer the transfer of one
of the dangling water molecules to the other dangling
water via a three--membered ring
results in a degenerate  W3$_{1}$ isomer structure of \cf{H7O3+}. 
This 
rearrangement reaction
is accompanied by proton transfer
and features a reaction barrier
of about \SI{23}{\kilo\joule\per\mol} which
is well reproduced by the NNP.
A very similar reaction leads in the
larger protonated water tertramer to
isomerization from the Eigen to a Zundel-like
conformation.
This 
process is characterized by
a similar barrier height
as the water transfer in the trimer of
around \SI{23}{\kilo\joule\per\mol},
however leading to an 
energetically higher-lying \cf{H9O4+}
isomer in this case.
Again, the NNP and coupled cluster reference
provide essentially identical reaction profiles with
only very minor differences.

\subsection{Molecular Dynamics and Path Integral Simulations}
\label{ssec:md}
After having confirmed that 
also 
reaction pathways
which drive these complexes far away from important stationary-points structures
are well reproduced by the 
NNP, we finally 
validate
the performance of the network 
when used in computer simulations with classical and quantum nuclear motion, i.e.
for molecular dynamics and path integral simulation techniques.
We recall 
that the performance of the NNP for truly uncorrelated
configurations has been already validated based on the
independent training set that was build up during the automated
fitting process and, therefore, reflects vastly different conditions.
In practice, however, it is 
also important to relate this accuracy to the fluctuations
as generated during production simulations in order to 
explicitly validate the stability of the NNP in practical applications. 
For that purpose, the largest cluster, \cf{H9O4+},
was simulated 
for \SI{25}{\pico\second}
{\em via}
classical MD and path integral MD employing the NNP
at \SI{1.67}{\kelvin} (PIMD), \SI{300}{\kelvin} (MD and PIMD),
as well as \SI{600}{\kelvin} (MD).

The classical MD 
trajectories were
initialized at the Eigen conformer,
while the PIMD runs
were
started at the Zundel-c conformation to
probe different regions of the PES.
Note that in the high temperature classical
simulation a variety of rearrangements are
observed, leading in the end of the \SI{25}{\pico\second}
to a configuration close to the Ring isomer.
The computational details of these simulations
can be found in Sec.~\ref{sec:comp-det}.
Afterwards, the last \num{100} steps of these
trajectories were reevaluated with the
coupled cluster reference to assess
the quality of the NNP during the simulations.
The resulting energy profiles are depicted
in Fig.~\ref{fig:md_test}.
Overall, 
the energy fluctuations are well reproduced and 
very good agreement between
the coupled cluster reference and the
NNP is observed.
It can therefore be concluded
that the NNP is also able to reliably
describe protonated water
clusters during classical MD
and quantum PIMD simulations
at various conditions covering very low and high temperatures,
where configurations that fluctuate far away from the optimized
stationary--point structures and MEPs are encountered.

It is important to note that such
NNP simulations can be routinely performed
on a normal desktop machine in a few minutes.
Thus, this approach renders
coupled cluster accuracy accessible
for exhaustive exploration of protonated water
clusters with different sampling techniques
and at various conditions.

\section{Conclusions and Outlook}
\label{sec:conclusion}

In conclusion, a systematic and fully automated
procedure to 
efficiently parameterize 
potential energy surfaces for finite sized clusters 
employing high--dimensional
neural network potentials has been developed.
This NNP fitting procedure provides
convincing agreement with the highly accurate
reference electronic structure method, which is
coupled cluster theory 
up to perturbative triples 
in an essentially converged basis set when used in conjunction with the F12 
explicitly correlated wavefunction ansatz,
namely CCSD(T*)-F12a/aug-cc-pVTZ.
The flexibility of the underlying functional relation
provided by the neural network ansatz
allows 
one
to easily identify deficiencies
in the training set which can be used to systematically
improve the potential while 
keeping the required total number of computationally demanding reference calculations to a minimum. 
This enables fast, automated and accurate development of
NNPs and therefore overcomes the obstacles of traditional fitting approaches.

For the chosen protonated water clusters 
from the monomer to the tetramer,
not only 
configurations close to equilibrium structures as characterized by 
stationary points
and small--amplitude oscillations, 
but also non--equilibrium regions of the potential energy landscape 
are described with 
essentially 
the same accuracy
using the same NNP for the four clusters. 
This holds true also for rearrangement reactions 
of these clusters involving proton transfer
as well as other interconversion pathways.
In addition, the NNP is equally well
suited for the usage in classical molecular dynamics 
as well as for simulations including the
quantum nature of nuclei
from ultra-low to ambient to high temperature.
Overall, the presented procedure will open the door to study
many other intriguing systems at a converged description
of the interactions in a straight--forward manner.

\section*{Supporting Information}

See the Supporting Information for
the
details on the specific NNP formalism used, 
a full description of the specific architecture
together with all optimized parameters of the developed NNP,
as well as for complementary analyses of the learning
curves for the final data set.
In addition, all coupled cluster reference calculations
are provided as Supporting Information free of charge 
on the \href{https://doi.org/10.1021/acs.jctc.9b00805}{ACS Publications website}.

\begin{acknowledgments}
It gives us great pleasure to thank Harald Forbert, Fabien Brieuc and
Felix Uhl for helpful discussions.
This research is part of the Cluster of Excellence RESOLV
(EXC~1069, EXC~2033 project number~390677874) 
funded by the \textit{Deutsche Forschungsgemeinschaft}, DFG.
C.S. acknowledges partial financial support from the 
\textit{Studienstiftung des Deutschen Volkes} as well as from the
\textit{Verband der Chemischen Industrie}.
J.B. thanks 
the DFG
for a Heisenberg professorship (Be3264/11-2 project number~329898176).
The computational resources were provided by HPC@ZEMOS,
HPC-RESOLV, and BOVILAB@RUB. 
\end{acknowledgments}
%
%

%
\input{manuscript.bbl}
%\bibliography{bibliography.bib}
%
%
%
\end{document}

%% file: manuscript.bbl
%merlin.mbs aipnum4-1.bst 2010-07-25 4.21a (PWD, AO, DPC) hacked
%Control: key (0)
%Control: author (8) initials jnrlst
%Control: editor formatted (1) identically to author
%Control: production of article title (-1) disabled
%Control: page (0) single
%Control: year (1) truncated
%Control: production of eprint (0) enabled
%